\documentstyle[prb,aps]{revtex} \def\narrowtext{} \tighten \twocolumn
\begin{document}

\draft

\title{Absence of Persistent Magnetic Oscillations in Type-II Superconductors}
\author{M.R. Norman}
\address{Materials Science Division, Argonne National Laboratory,
Argonne, IL 60439}
\author{A.H. MacDonald}
\address{Department of Physics, Indiana University,
Bloomington, IN 47405}

\address{%
\begin{minipage}[t]{6.0in}
\begin{abstract}
We report on a numerical study intended to examine 
the possibility that magnetic oscillations persist in
type II superconductors beyond the point where the pairing self-energy 
exceeds the normal state Landau level separation.  Our work is based on 
the self-consistent numerical solution for model superconductors 
of the Bogoliubov-deGennes equations for the vortex lattice state.
In the regime where the pairing self-energy is smaller than
the cyclotron energy, magnetic
oscillations resulting from Landau level quantization are suppressed 
by the broadening of quasiparticle Landau levels
due to the non-uniform order parameter of the vortex lattice state,
and by splittings of the quasiparticle bands.
Plausible arguments that the latter effect can lead to a 
sign change of the fundamental harmonic of the magnetic oscillations 
when the pairing self-energy is comparable to the
cyclotron energy are shown to be flawed.  Our calculations indicate 
that magnetic oscillations are strongly suppressed once the 
pairing self-energy exceeds the Landau level separation.  
\typeout{polish abstract}
\end{abstract}
\end{minipage}}

\maketitle

\narrowtext

\section{Introduction}

de Haas-van Alphen (dHvA)
oscillations in the mixed state of type-II superconductors,
discovered\cite{nbse2} in $NbSe_2$ some time ago, have 
recently\cite{v3si,nb3sn,bedt} been observed in several additional materials. 
The oscillations are damped 
relative to those in the normal state and become unobservable at 
sufficiently weak external magnetic fields.  These findings
have led to a number of theoretical studies of the modification 
of normal state Landau level structure in the mixed state.
\cite{rasolt,dukan,stephen,maki,maniv,raja,miyake,miller}
Conclusions from these studies are not always completely consistent
and no widely accepted picture which covers
all regimes of magnetic field has emerged from this work.  
Recently we reported on a thorough numerical study of the quasiparticle 
band structure obtained by solving the Bogoliubov-de Gennes 
(BdG) mean-field equations in the vortex lattice state of a simple model
two-dimensional superconductor.\cite{dhva}
We found that at fields near $H_{c2}$, magnetic
oscillations were clearly present, but that these were rapidly 
damped as the superconducting self-energy strengthened at 
weaker fields.  We argued and found partial numerical 
support for the assertion that the effect of 
superconductivity was similar to the effect of a disorder 
broadening of the normal state Landau levels, proportional
to the pairing self-energy times $n_{\mu}^{-1/4}$ where
$n_{\mu}$ is the Landau level index at the Fermi level.
We also found that once the pairing self-energy became comparable
to the Landau level separation, the quasiparticle electronic structure
in the vortex lattice state entered a complicated crossover regime
which simplified with increasing pairing self-energy only
when unambiguous vortex cores with associated bound states emerged.
While magnetic oscillations were essentially absent once 
the vortex cores became distinct, we were unable to draw any clear 
conclusions concerning magnetic oscillations in the crossover regime.
These calculations did 
indicate the possibility of a phase shift of $\pi$ for magnetic
oscillations in the crossover regime, but the origin of
this phase shift was not understood.  Such a phase shift was also
found in earlier work by Maniv {\em et al}\cite{maniv1} based
on an expansion of the free
energy to fourth order in the order parameter.  Recently, Maniv
{\em et al.}\cite{maniv2} have attributed this phase shift to 
a splitting of Landau levels in the vortex lattice state 
which they associate with the two
vortices per electron magnetic flux quanta in the vortex lattice state.
This suggestion has motivated us to examine the crossover regime
in greater detail.    

Our study is based on numerical solution of the BdG
equations\cite{bdg} for a model superconductor with a BCS pairing interaction,
{\it i.e.} a $\delta$-function attractive interaction modified by
an energy cut-off.  We solve the BdG equations in a
Landau level basis so that the band energy quantization which is the
source of magnetic oscillations is incorporated in an exact way.
The formalism necessary to carry out these calculations in a convenient way is 
fully described in our earlier work\cite{physicac,dhva} and 
briefly summarized below.
This approach necessitates a number of practical limitations on
the scope of our study.
(i) Numerical problems which arise because of oscillations
in high Landau index quantum wavefunctions make it convenient to 
restrict our attention to single-particle states with 
Landau level indices smaller than $\approx 60$. (ii) We consider 
only two-dimensional electron systems; adding a third dimension
creates no formal difficulty but does add to an already
considerable computational burden.  
(iii) We approximate the magnetic field by its spatial average.
The most serious of these limitations 
is the restriction to moderately large Landau level indices.
Two-dimensional models will, if anything,
overestimate the importance of magnetic oscillations and are 
even appropriate for some systems of current interest.  The 
screening corrections to the uniform external magnetic 
field are small close to $H_{c2}$ and are approximately uniform
themselves, except for external fields close to $H_{c1}$.  

In the present study 
Zeeman splitting is ignored since its effects are well understood.
Most of the results we discuss use the grand canonical ensemble
rather than the canonical ensemble appropriate to experimental 
systems, since this eliminates the problem of determining the 
chemical potential self-consistently.  
In the normal state, there is little
difference between magnetic oscillations in canonical and grand
canonical ensembles for Landau level indices larger than
about six.\cite{shoen}  In the mixed state, however, canonical 
and grand canonical ensemble results may differ.
We have therefore, in some cases, executed the Legendre transform
from the grand canonical
to the canonical ensemble numerically in order to quantify the 
importance of magnetic oscillations in the chemical potential.

In Section II of this paper we summarize the BdG
formalism which is the basis of our numerical calculations.
In Section III we discuss the devolution of the Landau level 
structure in the quasiparticle spectrum as the superconducting
order strengthens.
It is this devolution which underlies the damping of 
dHvA oscillations in the mixed state.  We find that a 
picture in which the normal state Landau levels simply broaden
captures little of the process, hence   
the substantial difficulty in developing a simple
analytic theory for the influence of 
superconductivity on dHvA oscillations analogous 
to the simple and successful theory for the influence of 
disorder.  In this section we discuss a plausible approximation
which suggests that
magnetic oscillations in the vortex lattice state in the 
crossover field regime will differ by a sign from those 
in the normal state.  The possible sign change is associated
with a splitting in the density of quasiparticle states 
associated with each Landau level at fields below $H_{c2}$.
We explain the origin of this splitting and comment on the 
failure of the commonly used diagonal approximation for 
the quasiparticle spectrum.
In Section IV we carefully examine magnetic oscillations in this regime
and find that the sign change does not survive a more 
thorough analysis.
Instead, the fundamental harmonic of the magnetization is strongly damped.
The magnetization in this regime has substantial variation with field
but the indications from our numerical calculations is that the 
field dependence is aperiodic.  We conclude in Section V with a brief summary.

\section{Bogoliubov-de Gennes formalism}

In zero field BCS theory, the pairing self-energy couples only
single-particle states at wavevectors $\vec k$ and $-\vec k$.
The property that the coupled states have the same band energy 
is favorable for the formation of a condensate of electron pairs. 
In a magnetic field the 
loss of time-reversal invariance makes it impossible
to achieve this situation.
The center-of-mass momentum of a pair of electrons,
which is zero for condensate pairs at zero magnetic field, 
has quantum fluctuations in a magnetic field $ \sim  \hbar / \ell$
where $\ell = (\hbar c / e B)^{1/2}$ is the quantum magnetic length.
Associated quantum fluctuations in the momenta 
of the individual electrons contributing to the pair lead\cite{cooperon}
to pairing between electrons in different Landau levels 
and therefore with different single-particle energies.
It is this qualitative difference
which is responsible, from a microscopic point of view,
for the decrease of $T_c$ in a magnetic field.  
The well known dependence of $H_{c2}$ on field,
obtained from semiclassical theory or (near $T_{c0}$) from
Ginzburg-Landau theory, reflects in the microscopic theory
primarily contributions from pairing between electrons in different 
orbital Landau levels.\cite{ajp} 
It is not possible to
understand the modification of Landau levels by superconductivity,
even in the regime near $H_{c2}$, unless one includes these
`off-diagonal' terms.\cite{dhva}  

The BdG mean-field equations for a superconductor
in a constant magnetic field\cite{physicac,ajp} replace the $2 \times 2$
secular matrix of BCS theory at zero magnetic field by a
secular matrix of order $2N$
(where $N$ is the number of Landau levels within a pairing cut-off energy)
for each wavevector, $\vec{k}$, in the Brillouin zone of the vortex lattice.
The diagonal (normal) electron and hole blocks of the secular
matrix are diagonal in the 
Landau level basis with elements given by $\xi_n$ and $- \xi_n$ 
respectively where $\xi_n \equiv 
(n+1/2)\hbar\omega_c-\mu$.  Here $\omega_c=
eB/mc$ is the cyclotron frequency and $\mu$ is the chemical potential.
It is the simplicity of the diagonal block which makes such a basis convenient.
The off-diagonal (pairing) blocks have matrix elements\cite{physicac}
\begin{equation}
F_{\vec{k}NM} = {-\lambda \hbar \omega_c \over 2} \sum_j \chi_{M+N-j}(\vec{k})
              D_j^{MN}\Delta_j
\label{eq:1}
\end{equation}
with
\begin{equation}
\chi_j (\vec{k}) = \sum_t e^{i2k_x a_x t}e^{-i \pi t^2 /2}\chi_j (2k_y l^2
             +2ta_x )
\label{eq:2}
\end{equation}
\begin{equation}
\chi_j(Y) = ({1 \over 2^j j! \sqrt{2\pi}})^{1/2}e^{-Y^2/4l^2}
H_j ({Y \over \sqrt{2}l})
\label{eq:3}
\end{equation}
($H_j$ a Hermite polynomial) and
\begin{eqnarray}
D_j^{NM} = ({j!(N+M-j)!N!M! \over 2^{N+M}})^{1/2} \nonumber \\
           \sum_{m=0}^j {(-1)^{N-m} \over (j-m)!(N+m-j)!(M-m)!m!}
\label{eq:4}
\end{eqnarray}
In these equations $\lambda$ is the BCS coupling constant
so that $\lambda \hbar \omega_c = V/(2 \pi l^2)$
where V is the strength of the attractive interaction.
The vortex lattice primitive vectors
are $(0,a_y)$ and $(a_x,-a_y/2)$ with $a_xa_y=\pi l^2$
($a_x=\sqrt{3}a_y/2$ for a triangular lattice).
The sum over $j$ in
Eq. 1 is over the possible partitionings of
the total quantized kinetic energy of the pair,
$\hbar \omega_c (N+M+1)$, into contributions from the
pair center of mass motion, $\hbar \omega_c (j+1/2)$, and
the pair relative motion, $\hbar \omega_c (N+M-j+1/2)$,
with $(D_j^{NM})^2$ the probability that a pair of electrons in
Landau levels $N$ and $M$ will have center-of-mass kinetic energy
$\hbar \omega_c (j + 1/2)$.

In this formalism the order parameter in the vortex lattice state is 
paramaterized by a small set of numbers, $\Delta_j$, which 
should be determined by solving the 
BdG equations self consistently:\cite{physicac}
\begin{eqnarray}
\Delta_j = -\sum_{NM} D_j^{MN} \sum_{\vec{k}} {2a_x \over N_kl}
    \chi_{M+N-j}^*(\vec{k}) \nonumber \\
    \sum_{\mu} (1-2f_{\vec{k}}^{\mu}) u_{N\vec{k}}^{\mu} v_{M\vec{k}}^{*\mu}
\label{eq:5}
\end{eqnarray}
where $E_{\vec{k}}^{\mu}$ is the $\mu$'th positive eigenvalue of the secular
matrix, $(u_{N\vec{k}}^{\mu},v_{M\vec{k}}^{\mu})$ is the corresponding
eigenvector, and $f_{\vec{k}}^{\mu}$ is the Fermi function.  $N_k=L_xL_y/
(2\pi l^2)$ is the number of k points ($L_xL_y$ is the area of the system).
The Abrikosov solution\cite{alex} for the order parameter near 
$H_{c2}$ corresponds to a solution with only $\Delta_0 \ne 0$ and 
it is easy to verify that this solution is recovered in the 
appropriate limit.  For a
triangular flux lattice, the lowest energy solution has $\Delta_j$ real and
non-zero only for $j=6m$ where $m$ is an integer.

We determine the magnetization by numerically differentiating 
the appropriate thermodynamic potential with respect to magnetic
field.  The grand potential may be expressed in the following 
form which we use for our numerical calculations:\cite{physicac}
\begin{equation}
\Omega = \sum_{N} \xi_N N_N + E_P - TS
\label{eq:6}
\end{equation}
where the pairing self-energy is
\begin{equation}
E_P = -\lambda \hbar \omega_c {lN_k \over 4a_x} \sum_j |\Delta_j|^2
\label{eq:7}
\end{equation}
Here $N_N$ is the occupation number of Landau level $N$
\begin{equation}
N_{N} = \frac{2}{N_k}\sum_{\mu\vec{k}} f_{\vec{k}}^{\mu}|u_{N\vec{k}}^{\mu}|^2
 + (1-f_{\vec{k}}^{\mu})|v_{N\vec{k}}^{\mu}|^2
\label{eq:8}
\end{equation}
and $S$ is the entropy:
\begin{equation}
S = -\frac{2k_B}{N_k}\sum_{\mu\vec{k}} (1-f_{\vec{k}}^{\mu})
\ln(1-f_{\vec{k}}^{\mu}) + f_{\vec{k}}^{\mu}\ln
f_{\vec{k}}^{\mu}.
\label{eq:9}
\end{equation}
For canonical ensemble calculations we calculate the free energy 
$F$ over a range of electron densities from the grand potential calculated
over a range of chemical potentials by using 
\begin{equation}
F = \Omega + \mu\sum_N N_N.
\label{eq:10}
\end{equation}
where both $F$ and the density depend parametrically on $\mu$. 
The canonical ensemble magnetization is determined by numerically
differentiating $F$ with respect to field at fixed density.  
A portion of the discussion of our results is motivated by 
an equivalent alternate expression for $\Omega$ 
in terms of quasiparticle energies:\cite{bardeen,stephen,maniv2}
\begin{equation}
\Omega = -\frac{2k_BT}{N_k}\sum_{\mu \vec k} \ln[2\cosh(\frac{E^{\mu}_{\vec k}}
{2k_BT})] + \sum_n \xi_n + E_P
\label{eq:11}
\end{equation}
The last term here is a double counting correction for the pair interaction
energy.

The magnetization is determined by numerically differentiation:
$M(B) =  -\partial \Omega/\partial B$.  In
practice, we generate results as a function of $n_{\mu}\equiv \mu/
\hbar \omega_c-1/2$ and calculate energies per state in the Landau
level in units of $\hbar \omega_c$. 
Therefore, the derivative for the magnetization has two
terms, the first coming from differentiating an explicit dependence on B
(that is, $\Omega=\Omega_0 B^2$, with one
power of B coming from $\hbar \omega_c$ and the other from
the Landau level degeneracy factor), the second from the dependence of
$\Omega_0$ on $n_{\mu}$ which is determined numerically.
Note that we do not need to perform separate calculations to 
determine the density dependence of $\Omega$ mentioned above and the 
field dependence of $\Omega$ required for the magnetization.  
Similarly, in the canonical ensemble the magnetization can 
be expressed in terms of the derivative of the corresponding 
dimensionless free energy with respect to $N=\sum_N N_N$.

\section{Landau Level Devolution in the Mixed State}

We first analyze the secular matrix in the limit of small $\Delta_0$.
Our objective
here is to understand the behavior of the mixed state quaisparticle
bands over one period of the normal state magnetic oscillations.
Consider the case where $n_{\mu}=n$ ($n$ an integer).
For this case, we note that for each electron energy in the
upper diagonal block, there
will be a hole energy of the same value in the lower diagonal block.
For the Landau level at $\mu$, these two have the same index ($n$), 
otherwise, their indices are different ($n+m$ and $n-m$).  When the
order parameter is small, the strongest
mixing of a particle in Landau level $n+m$ will be with 
a hole in Landau level $n-m$.
The degeneracy of the particle and hole levels
will be lifted by the matrix elements in the pairing block 
which are, in general, off diagonal in Landau level index.
At a given $\vec k$ 
the two levels will be split by $2 |F_{n+m,n-m}|$ for all Landau levels
within the pairing cut-off.  In particular, one of the quasiparticle energy
levels at zero in the normal state will be shifted up by $|F_{nn}|$ while
one of the quasiparticle levels at $\hbar \omega_c$
in the normal state will be shifted down by $|F_{n+1,n-1}|$.
Obviously, this splitting cannot continue to grow 
indefinitely since these two levels will eventually approach each other,
leading to an avoided crossing.
A similar degeneracy occurs when $n_{\mu}=n+1/2$
with the electron level at Landau level index $n+1+m$ and the 
hole level at Landau level index $n-m$ being degenerate 
resulting in a similar splitting of each Landau level.  
In this case, one of the two quasiparticle levels which 
has energy $1/2\hbar \omega_c$ in the normal state will be shifted
down by $|F_{n+1,n}|$.  For $n_{\mu}=n+1/4$ (or $n+3/4$),
the level repulsion effect is weak; that is, the Landau level splitting is most
pronounced when degeneracies occur in the normal state,
partially invalidating the analogy to Zeeman splitting 
suggested by Maniv {\em et al}.\cite{maniv2}

We illustrate these points in an
approximation where all matrix elements in the pairing blocks are taken to be
the same constant, $-1/2 (n_{\mu}\pi)^{-1/4} \Delta_0 \hbar \omega_c$ which
is the large $N$
limit of the matrix element at the chemical potential if $\lambda=1$ and
$\chi$ is set to unity.\cite{ajp}  For cases considered in this paper,
we take the cut-off, $\omega_D$, to be $1/2\mu$ (thus for $n_{\mu}=20$, Landau
levels 10-30 are involved in the pairing).
The resulting eigenvalues for the above three cases
are plotted as a function of $\Delta_0$ in Fig. 1.  In the large $\Delta_0$
limit, oscillations in the low-energy quasiparticle eigenvalue spectrum
are about the same magnitude and shifted by half a period
relative to the oscillations in the normal state.
The expression for the grand potential in terms of quasiparticle
energies suggests that this might lead to a $\pi$ phase shift
in the Fourier transform of the
magnetization relative to the normal state case, {\it i.e.} to a change
in sign of the oscillatory contribution to the grand potential.
We say ``might"
since it is not obvious, even from Eq. 11, that a phase shift in 
the oscillations of low-energy  quasiparticle energies will necessarily 
show up as a phase shift of the magnetization.
(Eq. 11 involves three terms and each contributes strongly to the
oscillatory dependence of $M(B)$.) 

To examine this idea in more detail we have solved the 
BdG self-consistently at several different $\lambda$ values
for $n_{\mu} \in (20,21)$ and calculated coefficients of 
the Fourier expansions of quantities of interest within this
interval.  In the normal state the Fourier expansion 
coefficients vary slowly with the Landau level index associated with
the interval over which the Fourier transform is performed, 
since the dominant variation with $n_{\mu}$ is  
periodic.  In the mixed state we will have to check
for this periodicity by verifying that the Fourier expansions
in successive intervals are similar.  We focus on the 
coefficient of the leading sine term in the Fourier expansion 
which is the dominant term in the normal state and refer to
the Fourier expansion coefficients as harmonics of the 
magnetic oscillation; the terminology 
anticipates a repetition of the same pattern in successive intervals
which does not always occur as we discuss in further detail 
below.   For the interval $n_{\mu} \in (20,21)$ 
we find that  
the zero of the fundamental sine harmonic
of the Fourier transform of $M(B)$
in this interval does indeed closely
correspond to the point where the three curves in Fig. 1 cross.
To test the degree of correspondence between the total oscillatory 
contribution to the grand potential and the contribution from the 
lowest band of quasiparticles,
we have also verified that the quantity
\begin{equation}
\tilde{E_1} = -\frac{k_BT}{N_k}\sum_{\vec k} \ln[2\cosh(\frac{E^1_{\vec k}}
{2k_BT})]
\label{eq:12}
\end{equation}
(with $1$ denoting the lowest quasiparticle band) has a magnetization whose
fundamental sine harmonic agrees quite closely with that of the total
magnetization.
Note that $\tilde{E_1}$ is essentially -1/2 the mean of the energies of
the lowest quasiparticle band suggesting that there is some validity in
associating magnetization oscillations with oscillations in the 
low-lying quaisparticle bands.  
(This similarity of leading harmonics occurs  
even though the shape of the two `magnetizations'
with respect to $n_{\mu}$ are quite different; the 
correspondence does not hold for higher harmonics).  Finally, we
again note the qualitative difference between Fig. 1 and what would be
expected if the splittings were simply proportional to $\Delta$ 
as in the Zeeman-splitting analogy proposed by Maniv {\em et al}.\cite{maniv2}
In this case, avoided crossing effects at larger $\Delta$ do not
occur and additional zeroes would occur in the harmonics at larger
$\Delta$. 

The behavior seen in Fig. 1 should be contrasted with the
commonly used diagonal approximation, where the only elements retained 
in the pairing blocks are diagonal in Landau level index.  In this case, the
eigenvalues are simply shifted from $\xi_N$ to $\sqrt{\xi_N^2 + |F_{NN}|^2}$.
In this approximation the level splitting effect occurs only
when $\xi_N =0$; otherwise all quasiparticle Landau levels are
shifted away from the Fermi level.  
Because of this qualitative failure, we do not feel that the diagonal
approximation is useful for understanding the electronic
structure of the vortex lattice state except 
for the Landau level closest to the Fermi level and then only
when $n=n_{\mu}$. 

To examine how Fig. 1 is changed when the constant 
matrix element approximation is abandoned and details of 
pairing in the vortex lattice state are properly accounted for,
we have solved Eqs. 1-4 as a function of $\Delta_0$.
As discussed in our earlier work,\cite{dhva} the use of a sharp cut-off
when solving the secular matrix leads to spurious effects in $M(B)$ associated
with the ratio of the cut-off energy to the cyclotron energy.\cite{mark}
To eliminate this, we elect to use a smooth cut-off with the pairing
interaction between Landau levels $N$ and $M$ scaled by
$\sqrt{W_N W_M}$ where
\begin{equation}
W_N = 1.55 e^{-(\xi_N/0.5\omega_D)^4}
\label{eq:13}
\end{equation}
In Fig. 2 we show a plot of the density of states for $n_{\mu}=20$ and
$\lambda \Delta_0=1$.  Each quasiparticle Landau level, not
only the Landau level closest to the Fermi energy, is split
into two roughly symmetric subbands.  This splitting
is due to particle-hole mixing.
We have been unable to uncover a detailed connection between 
this splitting and the fact, emphasized by Maniv {\em et al.},
that two superconducting flux quanta pass through
each area of the vortex lattice state enclosing one electronic flux
quantum.  

In Fig. 3, we show results for the vortex lattice quasiparticle bands
which are analogous to those of Fig. 1
obtained using the constant matrix element approximation.
The plotted eigenvalue
in this case is the mean eigenvalue of the lowest band using a 66 $\vec{k}$
point grid
in the irreducible triangle (1/12) of the vortex lattice magnetic 
Brillouin zone.
The results look very similar to Fig. 1 up to the point where the curves
cross.  This crossing point is close to 
the point where 
the spatially averaged pairing self-energy in a coordinate
representation ($F_0 \simeq 0.44\lambda \Delta_0 \hbar \omega_c$) is equal
to $\hbar \omega_c$ ($F_0$ being the vortex lattice analogue of the BCS gap).
For smaller values of $\lambda \Delta_0$,
we are in the quantum regime
where we expect strong magnetic oscillations.
As $\lambda \Delta_0$ increases the dependence of the eigenvalues on
$n_{\mu}$ weakens and magnetic oscillations are correspondingly 
damped.  The oscillations are further damped in this regime
by the non-zero width of the Landau levels which reflects the 
the non-uniformity of the order parameter.
The width is linear in $\Delta_0$ for small $\Delta_0$ and 
should
lead to an exponential suppression of magnetic oscillations with an
effective scattering rate linear in $\Delta_0$.\cite{dhva}
At higher values of $\lambda \Delta_0$ we initially enter into
the crossover regime and then into the regime where 
well-defined vortex cores emerge.  
The fact that the mean eigenvalues increase in this regime 
reflects the crossover of the lowest energy quasiparticle states to 
vortex-core bound states.\cite{dhva}   The eigenvalues clearly 
still have a substantial dependence on $n_{\mu}$ within the 
interval $(20,21)$, at least in the crossover regime, although
the dependence is much weaker than in the constant matrix element
approximation.

Up to this point we have been performing calculations at fixed 
$ \lambda \Delta_0$, {\it i.e.} at fixed pairing self-energy.
To compute the magnetization we should in principle determine
$\Delta_0$ self-consistently at each value of $n_{\mu}$ and 
keep $\lambda$ fixed.  To facilitate comparisons with
the preceeding results for the quasiparticle bands we have
chosen instead to allow $\lambda$ to vary with $n_{\mu}$ 
so that self-consistency is achieved at a fixed value 
of $\lambda \Delta_0$.  This self-consistent value of $\lambda$ at 
a fixed $\lambda \Delta_0$ ($\tilde \lambda$) 
is easily determined\cite{dhva} by using Eq.(5) to calculate 
the output value ($\Delta_0^{out}$) at $\lambda =1$:
$\tilde{\lambda} \equiv \lambda \Delta_0^{in}/\Delta_0^{out}$.
Results are shown in Fig. 4 for the fundamental sine harmonic of the Fourier
transforms of $M(B)$ and $\tilde{E_1}$ versus $\lambda \Delta_0$.
A zero in the harmonic of $M(B)$ occurs for $F_0 \sim 1.6\hbar \omega_c$
(similar results are found for self-consistent calculations at fixed
$\lambda$).  The zero of the harmonic of $\tilde{E_1}$ is close to
the zero for $M(B)$ as claimed earlier.  
We note that in the small $\Delta_0$ regime, the dependence of the harmonic
on $\Delta_0$
contains both linear and quadratic terms.  The calculations are consistent
with a crossover from quadratic to linear behavior when the quantity
$F_0/n_{\mu}^{1/4}$ exceeds $2\pi k_BT$.
We also present in Fig. 4 canonical ensemble results generated from the grand
canonical calculations by a Legendre transform. 
Deviations from the grand canonical ensemble 
results occur at small $\Delta_0$.\cite{discont}
The important point, though, is that
the zeros of the harmonics in the two schemes agree.

\section{Absence of Persistent Magnetic Oscillations}

The calculations in the previous section discussed the 
variation of different properties of the vortex lattice state
within one particular period ($n_{\mu} \in  (20,21)$) of 
the normal state magnetic oscillations.  In order for the 
magnetic oscillations to persist in the vortex lattice state,
the same pattern of variation must occur over many
periods of the normal state magnetic oscillations.  To 
investigate whether or not this is the case we have  
studied the dependence of superconducting properties on 
field through a number of periods of the normal state oscillations.
The small $\Delta_0$ behavior always
involves the quantity $\Delta_0/n_{\mu}^{1/4}$ and retains the normal state 
magnetic oscillations with reduced amplitude. 
The zero and subsequent sign reversal of the fundamental harmonic with
increasing $\Delta_0$, however,
does not occur in every normal state oscillation period.
In Fig. 5, we plot the sine
of the fundamental harmonic versus $n_{\mu}$ for a value of $\lambda \Delta_0$
equal to 4.75 (where the weak maximum in Fig. 4 occurs).  These results
show that no clear component of the magnetization with the normal
state period survives in the crossover regime.  The harmonic
of the Fourier tranform of the magnetization in the finite interval
from $n_{\mu}$ to $n_{\mu}+1$  in this regime varies in sign and 
magnitude with no pattern we have been able to discern, consistent with
results presented in our earlier work.\cite{dhva}
Corresponding variations occur in the lowest band quasiparticle energies.
In Fig. 6 we plot the eigenvalue means as in Fig. 3 but
for the case $n_{\mu} \in (24,25)$.  One sees that the three curves converge
together as in Fig. 3 but this time do not cross when the crossover
regime is entered.  That is, the crossing effect of Fig. 3 may or may not
occur depending on Landau level index.
To emphasize this, we plot in Fig. 7 the
fundamental harmonic averaged over two different six period intervals.
We see that the phase shift effect of Fig. 4 has been completely washed out
and the fundamental harmonic is smoothly damped to zero as $F_0$
increases beyond $\hbar \omega_c$.

\section{Concluding Remarks} 

Experimental evaluations of dHvA
oscillation amplitudes are based on Fourier transforms over many
periods of oscillations.  The results in the proceeding section
indicate that no measureable oscillation with the 
normal state period or with any other period we have been able 
to recognize occurs once the typical value of the pairing self-energy
becomes comparable to the Landau level separation.  Because we work
with relatively small Landau level indices compared with the 
typical experimental situation, we are not able to completely eliminate
the possibility that oscillations in this regime are periodic 
with a different periodicity or with a periodicity
in $B$ rather\cite{weissoscillation} than in $B^{-1}$, 
although we have looked for such patterns without success and 
are reasonably confident that they do not exist.
It seems clear that in the 
3D case where many Landau levels contribute even for a fixed
field, that magnetic oscillations will
be even more strongly suppressed.
Disorder broadening, which we have neglected,
will damp the oscillatory signal beyond that 
calculated here.

In conclusion, we have done a detailed analysis of the nature of the
quasiparticle states in the field regime near the upper critical field of a 2D
type-II superconductor.  We find that for small $\Delta$, {\it all} 
Landau levels and not just the Landau level at the Fermi energy are split.
This property is associated with the absence of time-reversal
symmetry in the presence of a magnetic field.
The splitting would be naively expected to 
lead to a sign change in the fundamental harmonic of the
Fourier transform of the magnetization for a value of the pairing self-energy
of order the cyclotron energy, analogous to the sign changes which 
can occur due to the spin-splitting of Landau levels.
However our numerical calculations show
that once the pairing self-energy is comparable to the normal state
Landau level separation,
although the spectrum of quasiparticle excitations and 
the magnetization have sizable variations on the magnetic
field scale of the normal state dHvA oscillations, the
variations are aperiodic.  Accordingly, we find that 
dHvA oscillations are strongly suppressed once 
this regime is reached.

\acknowledgments

This work was supported in part by the U.S. Dept. of Energy,
Basic Energy Sciences, under Contract No. W-31-109-ENG-38,
and in part by the National Science Foundation under grant
DMR-9416906.  
The original version of the computer code used for these calculations
was written by Hiroshi Akera.  The authors thank  
Tsofar Maniv for stimulating interactions and
Steven Hayden for some discussions about experimental data.

\begin{figure}
\caption{Lowest eigenvalue ($\hbar \omega_c$ units) vs $\lambda \Delta_0$
in an approximation where all pairing matrix elements are the same constant
[$-1/2(n_{\mu}\pi)^{-1/4}\Delta_0\hbar\omega_c$] for $n_{\mu}=20$
(solid points), $n_{\mu}=20.25$ (pluses), and $n_{\mu}=20.5$ (open points).}
\label{fig1}
\end{figure}

\begin{figure}
\caption{Density of states versus energy ($\hbar \omega_c$ units) for
$\lambda \Delta_0=1$ and $n_{\mu}=20$.}
\label{fig2}
\end{figure}

\begin{figure}
\caption{Mean eigenvalue ($\hbar \omega_c$
units) of the lowest band vs $\lambda \Delta_0$ for the flux lattice.
Same notation as Fig. 1.}
\label{fig3}
\end{figure}

\begin{figure}
\caption{Fundamental sine harmonic of $M(B)$ (solid points) and of the
magnetization of -1/2 the
mean of the first quasiparticle band (open points) vs $\lambda \Delta_0$
(grand canonical) from the
period $n_{\mu} \in (20,21)$.  The pluses
are results for $M(B)$ generated in the canonical ensemble.}
\label{fig4}
\end{figure}

\begin{figure}
\caption{Fundamental sine harmonic of $M(B)$ vs $n_{\mu}$ for $\lambda\Delta_0$
=4.75.  Each point represents a calculation over a single period.}
\label{fig5}
\end{figure}

\begin{figure}
\caption{Mean eigenvalue ($\hbar\omega_c$ units) of the lowest band vs
$\lambda\Delta_0$ for $n_{\mu}=24$
(solid points), $n_{\mu}=24.25$ (pluses), and $n_{\mu}=24.5$ (open points).}
\label{fig6}
\end{figure}

\begin{figure}
\caption{Fundamental sine harmonic of $M(B)$ vs $\lambda\Delta_0$ for the
periods $n_{\mu} \in (20,26)$ (solid points) and $n_{\mu} \in (21,27)$
(open points).}
\label{fig7}
\end{figure}


\begin{references}
\bibitem{discont} Since the canonical ensemble and grand canonical ensemble
results agree in the normal state, Fig. 4 implies a discontinuity in the
canonical harmonic as $\Delta_0$ approaches zero.  We have verified that 
the harmonic rapidly crosses over to the normal state result for small
$\Delta_0$, with the crossover becoming sharper as the temperature is lowered.
This effect was not seen for fixed $\lambda$, where we found that the two
schemes always agreed quite closely.
\end{references}
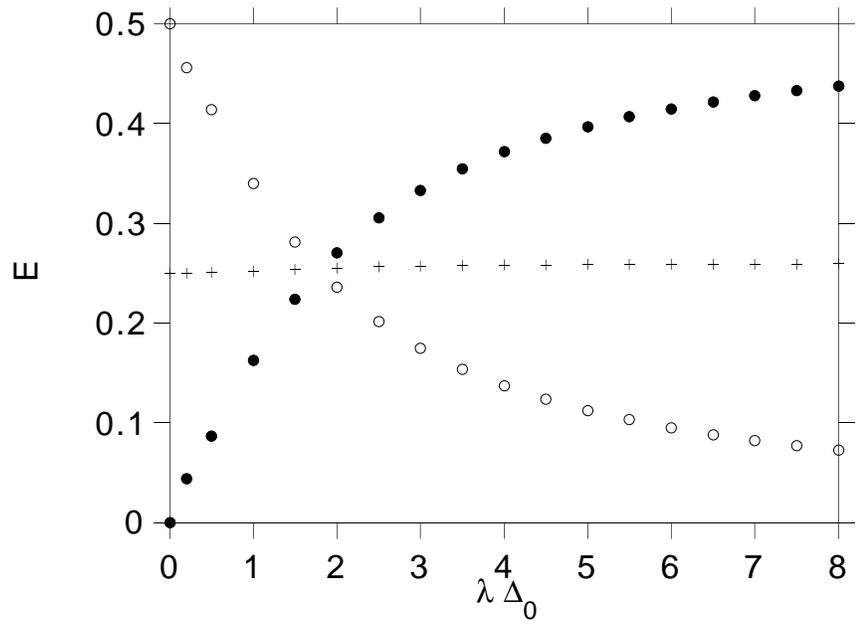
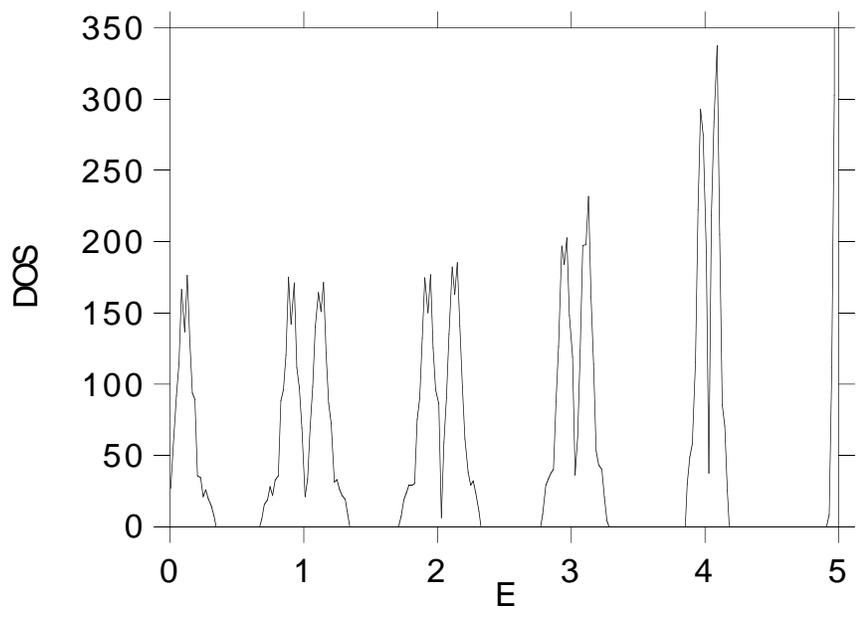
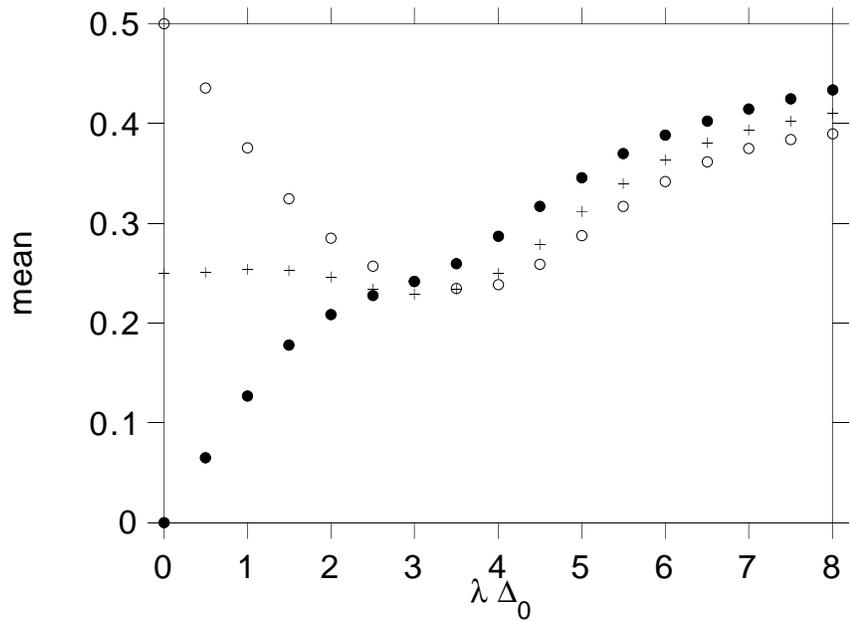
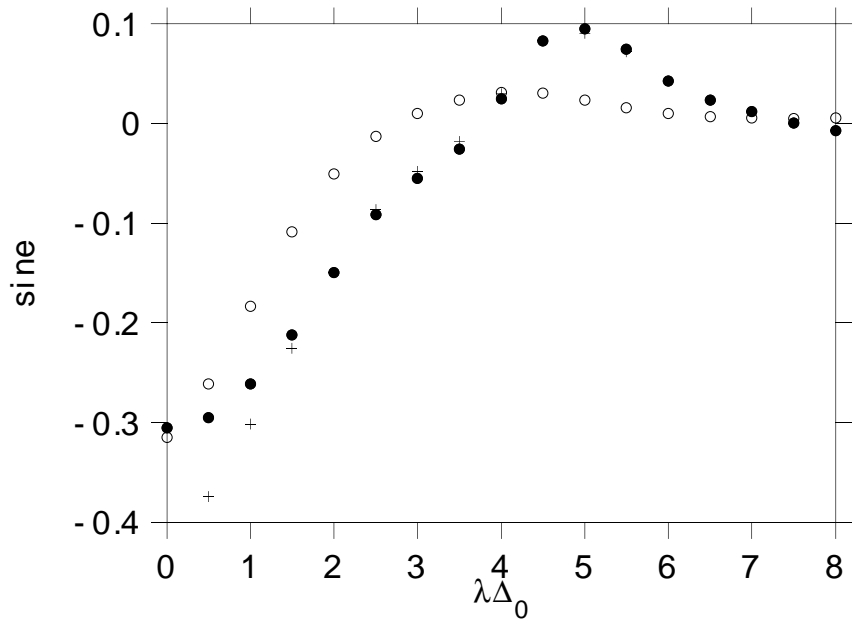
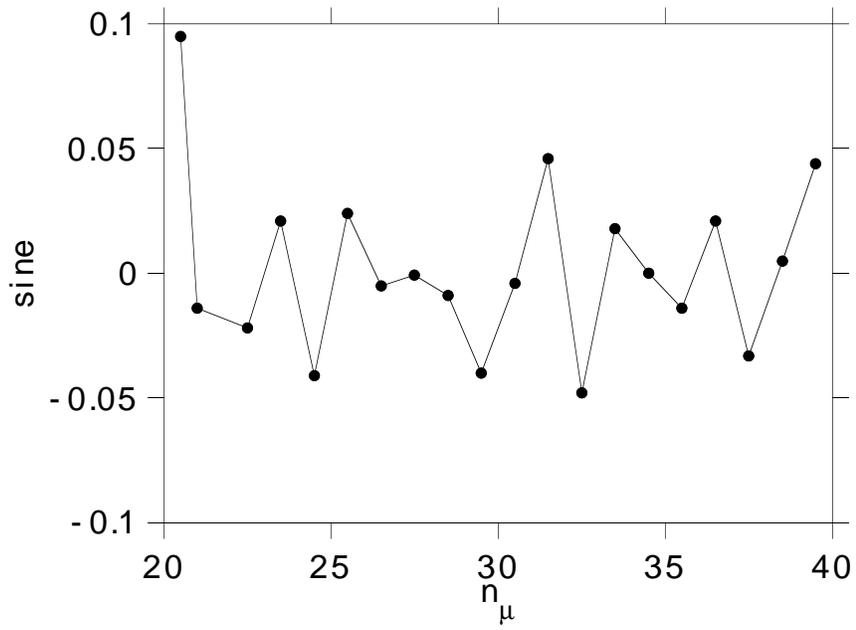
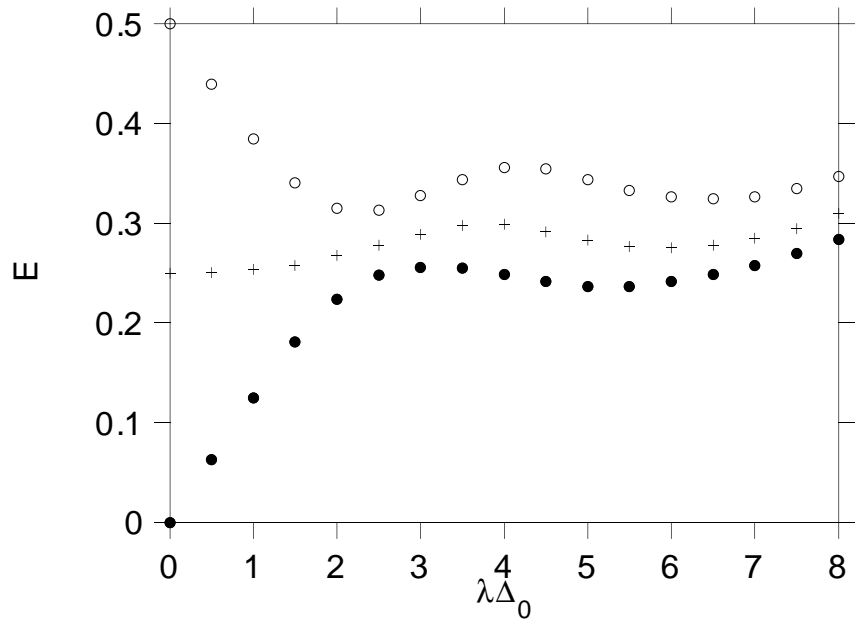
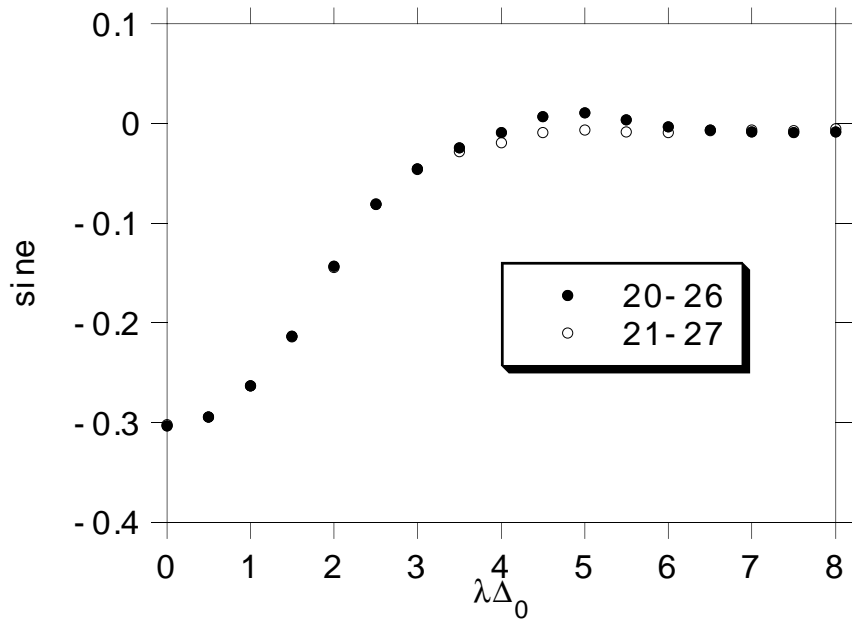
\end{document}